\documentclass{acm_proc_article-sp}
\usepackage{graphicx}
\usepackage{tabularx}
\usepackage{booktabs}
\usepackage{multirow}
\usepackage{dblfloatfix}
\begin{document}

\conferenceinfo{}{Bloomberg Data for Good Exchange 2017, NY, USA}

\title{Equity in 311 Reporting: Understanding Socio-Spatial Differentials in the Propensity to Complain}

\numberofauthors{3}
\author{
\alignauthor
Constantine Kontokosta\\
       \affaddr{Asst. Prof. (Urban Informatics)\\New York University}\\
       \email{ckontokosta@nyu.edu}
\alignauthor
Boyeong Hong\\
       \affaddr{Civic Analytics Fellow\\New York University (CUSP)}\\
\email{boyeong.hong@nyu.edu}
\vspace{2mm} 
\and
\alignauthor
Kristi Korsberg\\
       \affaddr{Civic Analytics Graduate Fellow\\New York University (CUSP)}\\
	\email{kk3374@nyu.edu}
\vspace{4mm} 
}

\maketitle
\begin{abstract}
\vspace{2mm} 
Cities across the United States are implementing information communication technologies in an effort to improve government services.  One such innovation in e-government is the creation of 311 systems, offering a centralized platform where citizens can request services, report non-emergency concerns, and obtain information about the city via hotline, mobile, or web-based applications. The NYC 311 service request system represents one of the most significant links between citizens and city government, accounting for more than 8,000,000 requests annually.  These systems are generating massive amounts of data that, when properly managed, cleaned, and mined, can yield significant insights into the real-time condition of the city. Increasingly, these data are being used to develop predictive models of citizen concerns and problem conditions within the city. However, predictive models trained on these data can suffer from biases in the propensity to make a request that can vary based on socio-economic and demographic characteristics of an area, cultural differences that can affect citizens' willingness to interact with their government, and differential access to Internet connectivity.  Using more than 20,000,000 311 requests - together with building violation data from the NYC Department of Buildings and the NYC Department of Housing Preservation and Development; property data from NYC Department of City Planning; and demographic and socioeconomic data from the U.S. Census American Community Survey - we develop a two-step methodology to evaluate the propensity to complain: (1) we predict, using a gradient boosting regression model, the likelihood of heating and hot water violations for a given building, and (2) we then compare the actual complaint volume for buildings with predicted violations to quantify discrepancies across the City. Our model predicting service request volumes over time will contribute to the efficiency of the 311 system by informing short- and long-term resource allocation strategy and improving the agency's performance in responding to requests. For instance, the outcome of our longitudinal pattern analysis allows the city to predict building safety hazards early and take action, leading to anticipatory safety and inspection actions. Furthermore, findings will provide novel insight into equity and community engagement through 311, and provide the basis for acknowledging and accounting for bias in machine learning applications trained on 311 data.
\end{abstract}


\keywords{E-government; complaint; propensity; civic engagement; data bias}

\section{Introduction}

 \begin{table*}[!b]
 \centering
 \caption{Description of Dataset used}
 \end{table*}
 \begin{figure*}[!b]
  	\centering
	\includegraphics[width=1.0\textwidth]{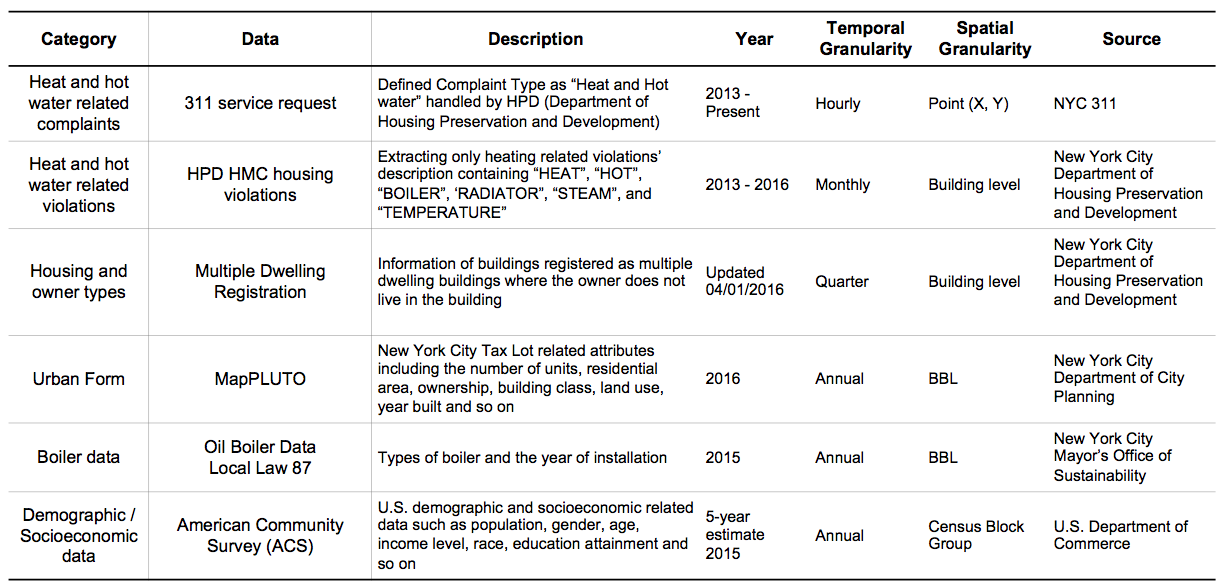}
 \end{figure*}

\vspace{2mm} E-government - the use of information communication technologies to improve public services - plays an increasingly important role in cities across the United States \cite{McClure2000}, \cite{Karen2001}. The most widely used are 311 complaint and service request systems, which are a critical mechanism for government-citizen interaction in cities such as New York City, Chicago, and Washington D.C. \cite{Minkoff2016}. These systems typically provide 24/7 customer service to residents and visitors by collecting reports of non-emergency concerns and information requests through a centralized platform, and then disseminating such reports to relevant city agencies \cite{NYC311}. The NYC 311 system is a large-scale data management enterprise; it receives more than 8,000,000 reports annually via telephone hotline, mobile app, or social media and tracks service response through real-time measurement and analysis of request activity \cite{Zha2014}, \cite{NYCEgovernment}. Twenty percent of all calls and website visits result in creation of 311 Service Requests\cite{NYCEgovernment}. These data provide an important picture of citizen concerns and problematic conditions across the city. More recently, these data are being used to predict problems and inform city decisions as a result the growing interest in the application of machine learning and data science to city management\cite{Wang2016}. For example, widely-available 311 data provide an opportunity to develop spatio-temporal proxy measures of neighborhood socioeconomic characteristics \cite{Wang2016}. Specific complaint data, coupled with other city records, have also been used to predict the emergence of unsafe or unhealthy conditions, including rodent infestations and illegally converted buildings.

Despite the positive potential of these data to improve quality-of-life, there is a real concern that analyses and predictive models may be biased without accounting for factors that could affect the likelihood of using 311 systems across demographic, socioeconomic, and cultural groups. Such disparities could result from a differing willingness to interact with government, varying levels of trust of government agencies, or lack of access to Internet connectivity. When using data that are not representative of the population, predictive analytics will be biased in their output, over- or under-estimating the location, severity, or progression of problematic city conditions, thus resulting in a mis-allocation of city services.

The aim of this research is to analyze the variance in the propensity to use the 311 system and to understand the relationship between socioeconomic, demographic, and cultural factors and complaint behavior. Quantifying the variability in the propensity to use 311 has the potential to affect resource allocation and strategy at NYC311 and other agencies that must respond to service requests. Furthermore, this research provides the basis for more equitably and accurately monitoring neighborhood conditions and changes in resident concerns over time. This paper is organized as follows: section two describes our data and methodology, section three introduces our preliminary findings, and section four presents a discussion of our results and future research. 

\section{Method}
\vspace{2mm}  
We develop a three-step approach to quantify the propensity to report heat and hot water problem through the 311 system. First, a a Gradient Boosting Classifier model is used to predict the likelihood of heat and hot water problems at the building-level across the City in a given heating season. Second, we compare the temporal and spatial patterns of heat and hot water complaints reported through 311 in a given heating season to the predicted likelihood of a heat and hot water violation at the particular building. Last, we contextualize the sociospatial pattern by identifying neighborhood characteristics that correlate with different observed complaint behavior patterns. 

In order to demonstrate the variability in the propensity to use 311 in NYC, we need to identify "ground-truth" - an objective measure of city conditions that is not reliant on citizen complaints to reveal. This is more challenging than it seems, and we develop an approach focused on heat and hot water complaints, a condition with significant implications for public health and safety, and one of the largest single 311 complaint categories. We narrow our scope to only include residential buildings that are registered as multiple dwellings where the owner does not live in the building (designated by the NYC Department Housing Preservation and Development), since we assume that one- and two-family housing with an owner present will address heating issues differently than larger, multi-tenant buildings. We focus on the heating season between October 1st and May 31st, which is defined as the period when building owners are required by law to provide heat throughout the building. According to our descriptive analysis, approximately 90\% of the City's heat and hot water complaints during the heating season are reported from multiple dwellings, such as walk-up and elevator apartment types. 

We then compare the predicted likelihood of heat and hot water violations based on building physical conditions in the 2016-2017 heating season to the actual heat and hot water complaints at the building-level. We hypothesize that four classification cases are detected; "matched" buildings where (1) the model predicts problems and where there are heat and hot water complaints or (2) the model does not predict problems and there are no heat and hot water complaints, and "mismatched" buildings where (3) the model indicates there should be an issue, but no complaints are reported or (4) the model indicates there should not be an issue, but complaints are reported. Based on these classified building types, we identify the locations where over- or under-reporting occurs and statistically compare differences in key demographic and socioeconomic characteristics. 

\subsection{Data}

\begin{figure}[t!]
  \centering
  \includegraphics[width=0.5\textwidth, height=0.4\textwidth]{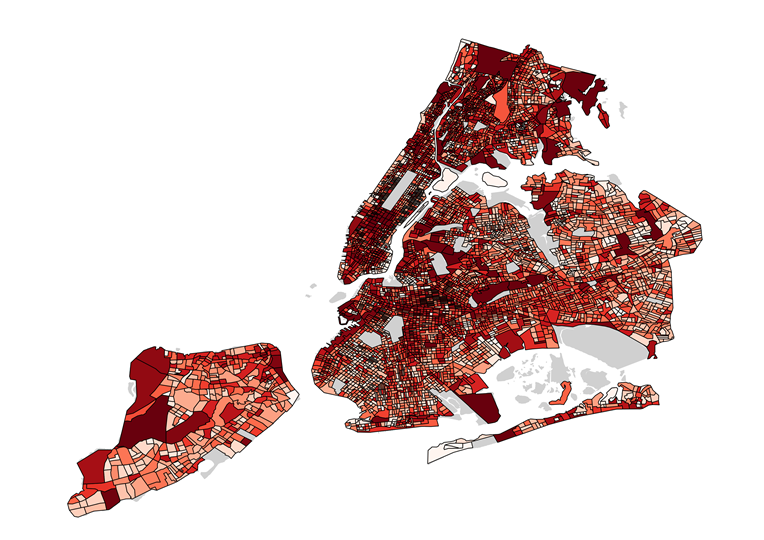}
  \caption{311 Service requests per capita at Census Block Group level in 2016 (Total requests: 2,150,628)}
\end{figure}

\vspace{2mm} For our analysis, we acquire and integrate a wide range of New York City datasets, described in Table 1. The major data source is 311 complaints data, which account for more than 2,000,000 service requests annually. Figure 1 shows the number of 311 complaints per person in 2016, providing a visualization of the residential population-normalized spatial pattern of 311 usage across the City.

Our analysis integrates several publicly available datasets: (1) complaints data pulled from 311 service request data, focusing on heat and hot water complaints handled by the NYC Department of Housing Preservation and Development (HPD); (2) building physical condition data integrated with heat and hot water-related violations issued by HPD, which includes property and building attribute data from NYC Department of City Planning (DCP), HPD, and the Mayor's Office of Sustainability (MOS); and (3) demographic and socioeconomic data from the U.S. Census American Community Survey. We geocode all datasets to the individual property (Borough-Block-Lot identifier) and use the Census Block Group as our areal unit for demographic and socioeconomic data. 

\subsection{Predictive model of building violations} \vspace{2mm}
The first step in our analysis is to predict the likelihood of heating-related violations based on building conditions and attributes, trained on data from the 2013-2016 heating seasons. The dependent variable is a binary variable for the presence of a building heat and hot water violation anytime during the heating seasons from 2013 to 2016, equal to 1 if a violation was issued over the study timeframe. Our predictors include physical building conditions, management structure, and systems data (Table 2). Over the study period, violations were issued for 5.14\% of the included multiple dwelling buildings, indicating that the dataset is imbalanced. We apply a Gradient Boosting Classifier with under-sampling of the majority class. After tuning parameters based on the confusion matrix, we compute the probability of heat and hot water violations at the building-level in 2016-2017 heating season (October 2016 ~ May 2017) and achieve a modest 75\% prediction accuracy. Briefly, property value, building shape, building age, ownership, building management, and physical relationship with the adjacent building are shown to be the most significant predictors.

\subsection{Classification by complaint behavior} \vspace{2mm}
We compare the predicted likelihood of potential problems derived from our model to a binary representation of the actual heat and hot water complaints reported through 311 during the 2016-2017 heating season. This is based on the number of complaints per residential building.
\begin{itemize}
	\item 0: No complaint reported (n=117,843)
    \item 1: At least one or more than one complaint reported (n=22,150)
\end{itemize}

\begin{table}[t!]
\caption{Input variables: building physical condition and property characteristics}
\smallskip
\def\arraystretch{1.5}
\begin{center}
\begin{tabularx}{\linewidth}{ c|X }
 \hline
 \textbf{Variable} & \textbf{Description}\\
 \hline
 Property value per sqft & Total property value / building area \\ \hline
 The number of units & Residential unit only \\ \hline
 Residential area per unit & average building area of unit \\ \hline
 Residential area ratio & Residential area / Total building area \\ \hline
 Building dimensions & Building width / depth \\ \hline
 Building age & - \\ \hline
 Basement code & Full or partial basement, no basement, unknown \\  \hline
 Proximity code & Physical relationship with adjacent buildings \\ \hline
 Ownership type & Individual, Corp , Company , others \\ \hline
 Super & whether building with super or not \\ \hline
 Boiler type & Gas, Oil, electricity, others \\ \hline
 Boiler age & - \\
 \hline
\end{tabularx}
\end{center}
\end{table}

\begin{table}[t!]
\caption{Building classification}
\smallskip
\def\arraystretch{2}
\begin{tabularx}{\linewidth}{X|l|X|X}
\hline
 &  & No predicted violation & Predicted violation\\
\hline
 &  & 0 & 1 \\
\hline
No reported 311 complaints & 0 & Type 1: No violation predicted, and no complaint reported & Type 2: Violation predicted, but no complaint reported \\
\hline
Reported 311 complaints & 1 & Type 3: No violation predicted, but complaints reported & Type 4: Violation predicted, and complaints reported \\
\hline
\end{tabularx}
\end{table}

Each multiple dwelling building is classified as one of the cases presented in Table 3, based on the prediction result and the binary representation of 311 complaints. We are particularly interested in investigating "mismatched" buildings: those that receive a high volume of complaints, yet no violation is predicted, and those buildings for which there are no complaints, but a violation is predicted. In order to analyze mismatched cases (under- and over-reporting), our building classification cases from Table 3 are re-grouped into the following:
\begin{itemize}
	\item "As expected" building cases (resident complaining as expected)\newline
- Type 1: No violation predicted, and no complaints reported \newline
- Type 4: Violation predicted, and complaints reported (regardless volume of complaints)\vspace{2mm} 
	\item "Mismatched" building cases (resident not complaining as expected)\newline
- Type 2: Violation predicted, but no complaints reported \newline
- Type 3: No violation predicted, but complaints reported (regardless volume of complaints) 
\end{itemize}

Mismatched cases are represented visually, with kernel density maps, in figures 2 and 3, which represent the complaint-violation pair for Type 2 and 3 buildings. Type 2 buildings are considered "under-reporting" and concentrated in upper Manhattan, the lower east side of Manhattan, and the Bushwick area in Brooklyn. Type 3 buildings are classified as "over-reporting" and most commonly found in the Upper East Side and Upper West Side of Manhattan, Midtown Manhattan, and parts of Brooklyn. These buildings (1) could have problems not captured in the predictive model of heating issues, and thus residents complain, (2) residents complain about issues that are not violating conditions, or (3) residents complain and the issue is fixed prior to the issuance of a violation. \vspace{2mm} 

\section{Findings} \vspace{2mm}

\begin{figure}[t!]
  \centering
  \includegraphics[width=0.5\textwidth]{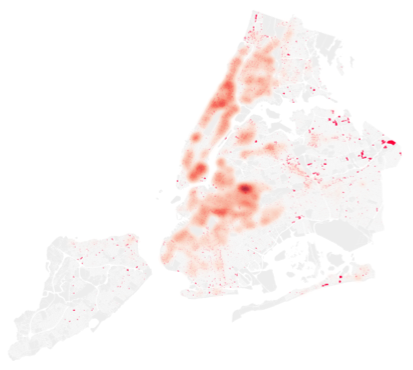}
  \caption{Hotspots of Type 2 buildings, 19,317 total buildings and 13.8\% of the sample}
\end{figure}
\begin{figure}[t!]  
  \centering
  \includegraphics[width=0.5\textwidth]{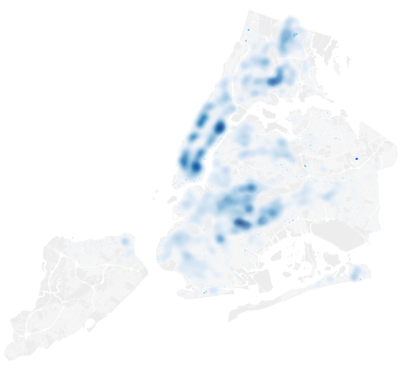}
  \caption{Hotspots of Type 3 buildings, 7,498 total buildings and 5.4\% of the sample}
\end{figure}

\begin{table}[h!]
\caption{Result of t-test comparing under-reporting case to over-reporting case}
\def\arraystretch{1.5}
\begin{tabularx}{\linewidth}{ X|c }
 \hline
 \textbf{Features} & \textbf{t-value}\\
 \hline
 Median rent & -14.42 \\
 Race diversity (How various races are mixed) & 13.81 \\
 Vacancy rate & -13.61 \\ 
 \% of minor population (Hispanics, Natives, etc.) & 12.57 \\ 
 Median income & -12.22 \\ 
 \% of English limited population & 9.74 \\ 
 \% of married population & -7.26 \\ 
 Unemployed rate & 5.94 \\
 \% of elderly population (over 70) & -5.09 \\
 \% of White & -4.85 \\
 \% of higher educated population (at least bachelor degree) & -3.06 \\
 \% of female & -2.12 \\
 \% of households living alone & -2.01 \\ \hline
 *** all p-value < .05 \\
 \hline
\end{tabularx}
\end{table}
Based on our results, we find clear discrepancies in complaint behavior after controlling for buildings of similar physical condition. Particularly, we are most interested in the comparison between the mismatched classification cases where complaint behavior deviates from expected (under- versus over-reporting population, shown in Figure 1 and 2).  

Differences in complaint behavior are observed across neighborhoods. We use t-tests to determine whether there are statistically significant differences between two mismatched groups. We conduct t-tests for gender, age, income, education level, race, language, and household characteristics. Table 4 shows the results of the t-tests. We observe statistically significant differences in certain demographic and socioeconomic characteristics between the under-reporting and over-reporting groups.

We find that neighborhoods that tend to under-report to 311 have a higher proportion of male residents, of unmarried population, and minority population; a higher unemployment rate; and more limited English speakers. This population appear to under-use (or not use at all) 311 even if they experience heat and hot water problems in their building.  Neighborhoods that tend to over-report, based on our analysis and assumptions, are those with higher rents and incomes and a higher proportion of female, elderly, and non-Hispanic White and Asian residents, with higher educational attainment. Based on these results, we find that socioeconomic status, household characteristics, and language proficiency have a non-trivial effect on the propensity to use 311 across the city.

\section{Discussion and Conclusion}
\vspace{2mm} Explaining the unequal use of NYC's 311 system is not without its challenges. The absence of "ground-truth" data complicates our ability to assess objective versus subjective measures of conditions. Also, it is hard to consider non-quantified characteristics, such as landlord response quality or building maintenance, which may impact the need to report a problem to 311. However, despite the limitations of our predictive modeling approach, the results presented here demonstrate that the propensity to complain through the 311 system in New York City does vary depending on residents' characteristics and their neighborhoods. This serves as initial evidence towards understanding the  propensity of different groups to complain, and to quantify the potential representativeness bias in 311 data. 

Our findings are applicable to a wide range of city operational decisions, such as resource allocation and preemptive inspections and maintenance. Overall 311 response efficiency can be improved if 311 works with related agencies to prioritize response to specific complaints that may be indicative of public safety hazards. For instance, a high-rate of complaints from a neighborhood classified as "under-reporting" may signal a more pressing problem than a similar complaint rate in "over-reporting" communities. As importantly, city agencies can encourage equity and community engagement through directed outreach and education in communities identified in our analysis. Future work will explore additional avenues for "ground-truth" observations and expand the analysis of the causes of varying complaint behavior.

\section{Acknowledgments}
Our thanks to the New York City Department of Information Technology and Telecommunications and NYC311 for providing data, insight, and feedback. This work has been supported, in part, by the John D. and Catherine T. MacArthur Foundation. Any opinions, findings, and conclusions expressed in this paper are those of the authors and do not necessarily reflect the views of any supporting institution. All errors remain the authors.



\nocite{*}
\bibliographystyle{abbrv}
\bibliography{references}

\end{document}